# Acoustic Metacages for Omnidirectional Sound Shielding

Chen Shen[1,2,*], Yangbo Xie[1], Junfei Li[1], Steven A. Cummer[1], and Yun Jing[2,†]

[1]Department of Electrical and Computer Engineering, Duke University, Durham, North Carolina 27708, USA

[2]Department of Mechanical and Aerospace Engineering, North Carolina State University, Raleigh, North Carolina 27695, USA

[*]chen.shen4@duke.edu

[†]yjing2@ncsu.edu

## Abstract

Conventional sound shielding structures typically prevent fluid transport between the exterior and interior. A design of a two-dimensional acoustic metacage with subwavelength thickness which can shield acoustic waves from all directions while allowing steady fluid flow is presented in this paper. The structure is designed based on acoustic gradient-index metasurfaces composed of open channels and shunted Helmholtz resonators. The strong parallel momentum on the metacage surface rejects in-plane sound at an arbitrary angle of incidence which leads to low sound transmission through the metacage. The performance of the proposed metacage is verified by numerical simulations and measurements on a three-dimensional printed prototype. The acoustic metacage has potential applications in sound insulation where steady fluid flow is necessary or advantageous.

## I. INTRODUCTION

Noise shielding and mitigation have long been a central topic in the field of acoustics [1]. Traditional noise shielding materials and structures rely on sound absorption and reflection to prevent the transmission of sound across a boundary. These materials or structures, however, typically stop both acoustic wave transmission and steady fluid flow across the boundary [2]. This characteristic severely limits their applications under circumstances in which the exchange of air is necessary or advantageous, such as noise reduction in environments where ventilation requires that air should be able to flow freely. Consider the noise control of cooling fans (Fig. 1), in which the free circulation of air is imperative to allow heat transfer and dissipation. Noise mitigation materials and structures such as high areal density panels and micro-perforated panels with backing cavities [3] are therefore not suitable as they prevent air flow.

Recent progress in acoustic metamaterials and metasurfaces has opened up new possibilities in manipulating waves [4–16] for many applications, including noise control, and they have shown substantial potential for building sound insulation panels [17–21]. However, they have yet to be proven useful for designing noise-control acoustic enclosures, especially those with openings. Several approaches have been proposed to block sound while enabling transport of air flow [22,23]. Although the transmission loss of these designs are high, the structures are generally bulky and may not insulate noise in an omni-directional manner or form an effective acoustic enclosure, therefore hindering their applications for certain real-world problems, such as insulating noise from fans and compressors. In optics, the concept of metacage has been recently proposed and metacages have been numerically shown to be able to shield electromagnetic (EM) waves in order to protect objects from radiation [24,25]. While Mirzaei *et al*. proposed a metacage design based on nanowires [24], Qian *et al*. suggested that gradient metallic grating is more feasible for constructing the metacage [25]. However, the latter strategy was demonstrated using the effective medium theory and no explicit design was provided. Furthermore, the shielding effect of optic metacages has yet to be experimentally observed.

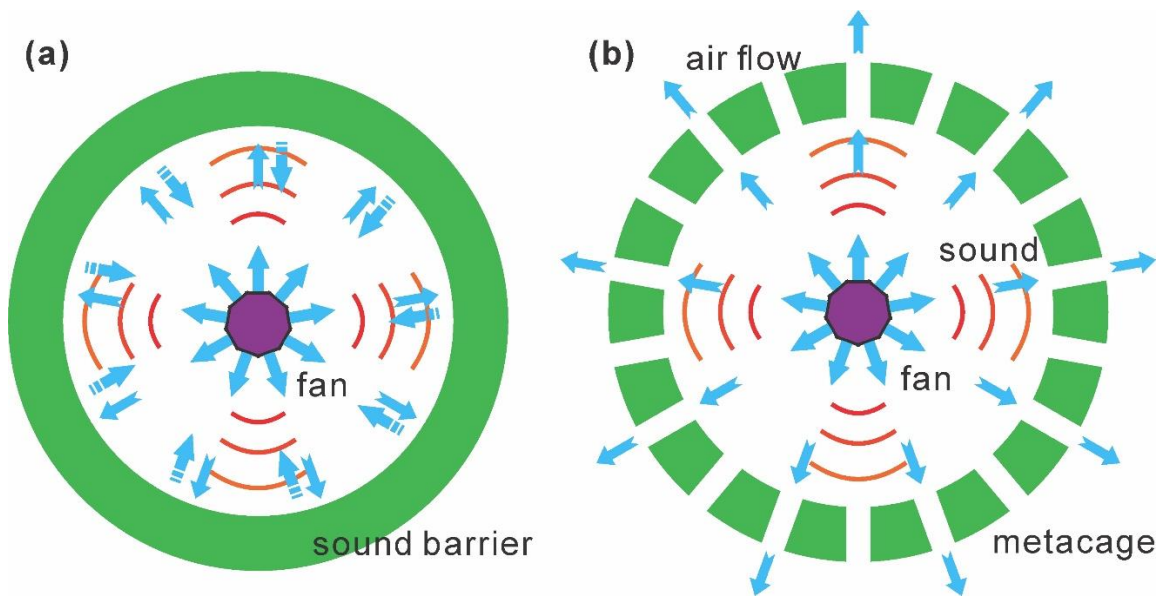

Fig. 1. Noise shielding using (a) a conventional material and (b) a new structure that allows exchange of air.

This paper investigates an acoustic metacage and presents a feasible design based on gradient-index metasurfaces (GIMs). Both simulation and experimental results demonstrate that the metacage is capable of shielding acoustic wave transmission from all angles, regardless whether the source is inside or outside the metacage. In the acoustic regime, GIMs have been reported to achieve anomalous reflection/refraction, controllable reflection, bottle beams, asymmetric transmission, among others [26–32]. In this paper, the sound transmission behavior of the GIM is first investigated by the means of mode coupling, which shows that incoming acoustic waves cannot be coupled into the transmission mode regardless of the angle of incidence when the phase gradient is sufficiently large. The GIM is further bent into a ring shape in order to create an acoustic metacage, though the shape can be arbitrary in theory as long as sufficient phase gradient is satisfied [33]. Shunted Helmholtz resonators with open channels are employed to construct the GIM and produce the required phase distribution. Numerical simulations are first carried out to verify the proposed acoustic metacage. A prototype is subsequently fabricated and validated experimentally.

## II. OMNIDIRECTIONAL SOUND SHIELDING USING GIM

First consider a GIM shown in Fig. 2(a). Without losing generality, the GIM is composed of four different unit cells in one period whose length is $d$. For an incoming wave with an angle of incidence $\theta_i$, the refraction angle $\theta_t$ can be calculated using the generalized Snell's law [26,34] which reads

$$(\sin\theta_t - \sin\theta_i)k_0 = \xi + nG \tag{1}$$

where $k_0$ is the wave number in free space, $\xi = \mathrm{d}\Phi/\mathrm{d}x$ is the phase gradient along the surface, $n$ is the order of diffraction, and $G = 2\pi/d$ is the reciprocal lattice vector. It is noted that the term $nG$ only appears when the period is comparable with the wavelength $\lambda$ at large angles of incidence [26]. For $0$th order diffraction, i.e., $n=0$, the critical angle for incoming waves to couple into propagating modes is expressed as $\theta_c = \sin^{-1}(1-\xi/k_0)$. When the period $d$ is a very small value, i.e, $d < \lambda/2$, we have

$\xi = 2\pi/d > 2k_0$. Subsequently, the critical angle $\theta_c$ becomes an imaginary number since $|1-\xi/k_0| > 1$, meaning that for an arbitrary angle of incidence $\theta_i$, the propagating mode is not allowed through the GIM when $n=0$.

On the other hand, for non-zero values of $n$, the transmission coefficients can be interpreted by a mode-coupling method [29,35,36]. Recall that $\xi = 2\pi/d$, the $y$ component of the wave vector of the $n$th order diffracted wave is given by $k_{y,n} = \sqrt{k_0^2 - \left[k_x + \frac{2\pi(n+1)}{d}\right]^2}$, where $k_x$ is the $x$-component of the incident wave vector. Since $d < \lambda/2$, for an arbitrary non-zero value of $n+1$ ($n$ is an integer), we shall have $\left|k_x + \frac{2\pi(n+1)}{d}\right| > \frac{2\pi}{\lambda} = k_0$ (note that $|k_x| < \frac{2\pi}{\lambda}$), indicating that $k_{y,n}$ becomes imaginary for any non-zero value of $n+1$. The transmitted waves are therefore evanescent and decay exponentially along the y-direction. It should be pointed out, however, that these waves can still travel in the x-direction and are essentially surface waves since $k_{x,n}$ ($k_x + \frac{2\pi(n+1)}{d}$) is still a real number. For $n = -1$, although the propagating waves are allowed, the transmission are extremely small due to destructive interference [25]. In other words, the overall transmission through the GIM for $d < \lambda/2$ is small regardless of the angle of incidence for any value of $n$. Consequently, such judiciously designed GIM can serve as an omnidirectional sound barrier for all-angle incoming waves.

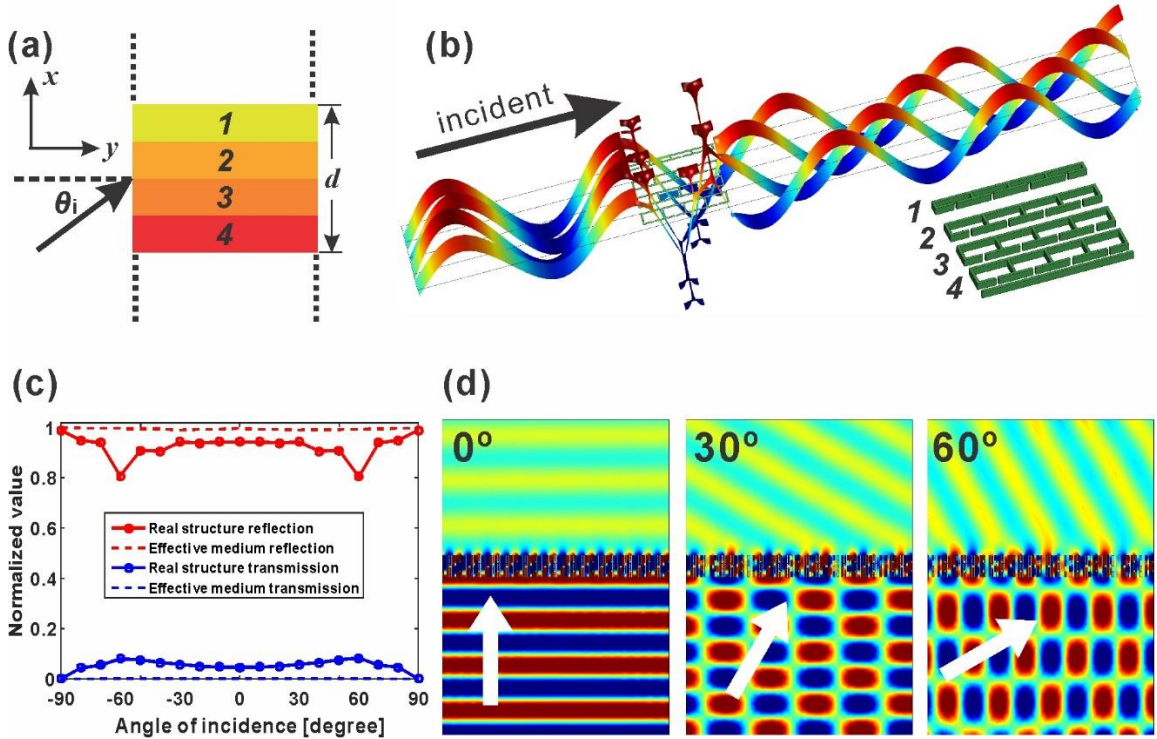

Fig. 2(a) Schematic of the GIM. (b) Simulation of transmission through the unit cells with phase shifts with a step of $\pi/2$. Inset shows the geometry of the units. (c) Normalized transmitted and reflected energy at different angles of incidence. (d) Acoustic pressure fields at three selected angles of incidence.

### III. PLANAR AND RING-SHAPED METASURFACE

The theory is verified through the case of a planar metasurface with the required large phase gradient. We design the unit cells using a hybrid structure consisting of open channels and shunted Helmholtz resonators although other existing metasurface unit cells in theory can be also used [26,32]. The original hybrid structure was proposed previously [30] and has shown an outstanding capability for controlling transmitted sound phase through the units [30,37,38]. Four individual unit cells are designed to maintain a uniform gradient of transmitted wave phase. Full-wave two-dimensional (2D) simulations are carried out by COMSOL Multiphysics v5.2, the Pressure Acoustics Module to verify the designed structure. Fig. 2(b) shows the acoustic fields propagating through the units, where high transmission can be observed when each unit cell is activated individually (i.e., no interaction between each two unit cells). However, when these unit cells form a metasurface and work collectively, omnidirectional sound reduction arises.

To demonstrate this, the transmission coefficients of acoustic waves at different angles of incidence are calculated numerically. The transmitted and reflected energy at a distance of 1.5 λ behind the metasurface is plotted as a function of incident angle as shown in Fig. 2(c). Overall, the metasurface effectively shields acoustic waves from arbitrary directions: the normalized energy transmission ($\sum I_t / \sum I_i$) calculated using the structure shown in Fig. 2(b) and using the effective medium are below 0.083 and 0.0016, respectively, at all angles, which correspond to 11dB and 28 dB transmission loss. The effective medium is characterized by effective refractive index and an impedance-matched condition so that there is an ideal interaction among the unit cells. The transmission coefficients of the real structure are much higher than those of the effective medium because the transmission can be sensitive to the variation of the phase and amplitude of the transmitted sound. Moreover, the impedance mismatch for real structures at oblique incidence may also contribute to the discrepancies between the real structure and effective medium simulations. Nevertheless, the simulation results of effective medium demonstrate the validity of the proposed structure for an omnidirectional sound barrier. As a reference, the acoustic fields of three cases where the incident angle is 0º, 30º and 60º are shown in Fig. 2(d). Most of the energy is reflected at the boundary and some surface acoustic waves can be observed along the top surface of the GIM, which agrees with the theoretical analysis.

To create an acoustic metacage that can reject acoustic transmission from all directions, we now bend the GIM into a ring shape. To ensure sufficient phase gradient along both outer and inner surfaces of the metacage, the unit cells are wedge-shaped, with each unit occupying a 5º segment of the circle as illustrated in Fig. 3(a). Four unit cells form a supercell, which is also the period of the metacage. The inner and outer radii of the metacage are 85 mm and 150 mm, respectively; the thickness of the metacage is 65 mm, which is about 0.47 λ at the designed operating frequency, i.e., 2.5 kHz. The design proposed in this study, however, can be scaled to work for any frequency of interest. The inner and outer widths of each supercell are 29.7 mm and 52.4 mm, respectively, both satisfying the condition $d < \lambda/2$ ($\lambda = 137.2$ mm). As discussed above, this condition ensures that the acoustic waves will be blocked from both interior and exterior of the metacage.

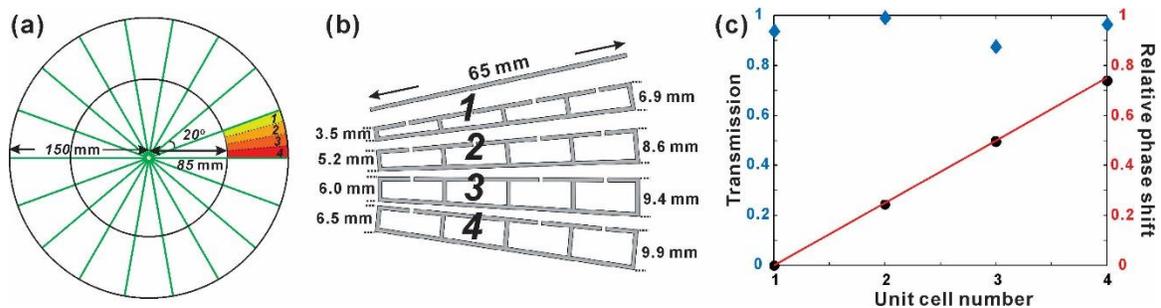

Fig. 3. (a) Schematic of the metacage. (b) Geometry of each unit cell. (c) Transmission and relative phase shift ($\Delta\phi / 2\pi$) through the unit cells. The red solid line represents the ideal phase shift of the unit cells.

The hybrid unit cell structure is modified and tailored to ensure that the accumulated phase change across adjacent unit cells has a phase difference of $\Delta\phi = \pi/2$ so that a supercell covers a complete $2\pi$ phase change. The heights of the Helmholtz resonators and the open channels gradually increase along the radial direction in order to be conformal with the ring shape of the metacage as illustrated in Fig. 3(b). The dimensions of the unit cells are highlighted in Fig. 3(b) and the average width of the open channels is 3.4 mm, allowing substantial airflow through the metacage. Figure 3(c) depicts the simulated phase shifts ($\Delta\phi/2\pi$) and normalized transmission coefficients of the unit cells. At 2.5 kHz, the transmission coefficients for all the unit cells are above 88% with accurate phase shift. The uniform transmission spectra ensures excellent coupling of the unit cells.

## IV. NUMERICAL AND EXPERIMENTAL INVESTIGATIONS

The performance of the proposed metacage is verified by both full wave simulations using real structures and measurements of a three-dimensional (3D) printed sample using acrylonitrile butadiene styrene (ABS) plastic whose density is 1230 kg/m$^3$ and speed of sound is 2230 m/s. The walls of the unit cells are assumed to be acoustically rigid due to the large impedance mismatch between the ABS and the background medium (air). We first study the case in which the metacage is exposed to a spatially-modulated Gaussian beam incident from the outside. The measurement where the metacage is exposed to a Gaussian beam is performed in a 2D waveguide [33]. Since the metacage has a curved geometry, it is illuminated by the Gaussian beam from various angles (i.e., normal incidence in the center and oblique incidence off the center). A fan (type FSY40S24M) is placed inside the metacage for the analysis of the effect of airflow. The air flow rate is 1.0 m/s at the inner surface. The sound transmission loss through the metacage with and without airflow is depicted in Fig. 4(a), where an average of more than 10 dB loss is observed within the frequency range from 2.2 to 2.6 kHz in the measurement. The resonance feature in simulation at around 2.4 kHz may have been caused by certain interaction among the unit cells such as Fano resonances [39] and is not observed in measurement due to fabrication tolerance and loss. The measured transmission loss is about the same with and without airflow, indicating that the metacage functions similarly with the existence of airflow. This is because the air flow rate (1.0 m/s) is much smaller than the sound speed in air (343 m/s) in our study. The airflow therefore has negligible effects on the acoustic properties of the metacage [40]. The high transmission loss in both simulation and measurement demonstrate the robustness of the metacage of shielding acoustic waves from all directions. The acoustic field behind the metacage is also scanned and compared with the simulation results shown in Figs. 4(b)-(d). As the metacage is almost axis-symmetrical, the results are similar for the Gaussian beam incident from other angles [33]. It can be seen that there is a low pressure "shadow" region behind the metacage, which is because the acoustic waves cannot penetrate the metacage. The different sizes of the shadow region in experiment and simulation might be caused by fabrication defect. In addition, since we did not consider the viscous loss through the unit cells, the induced dispersion may also lead to imperfect phase modulation [33,41].

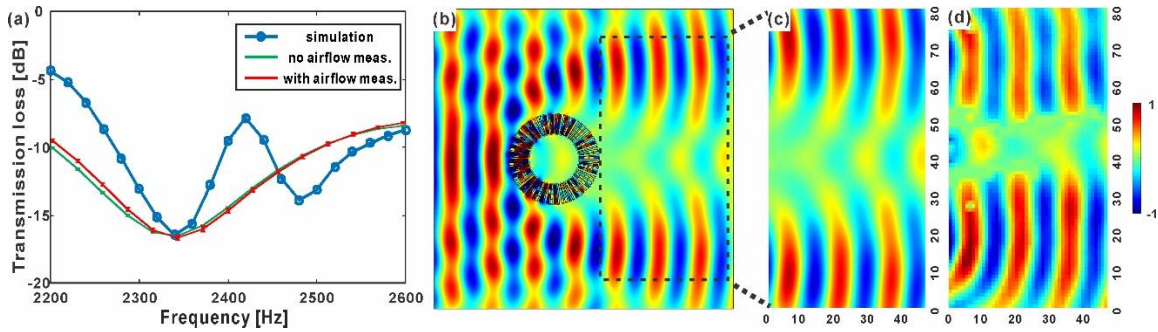

Fig. 4. Outside-to-inside performance of the proposed metacage. (a) Sound transmission loss through the metacage. The error bars are included for the measurement results and are computed out of four measurements. (b) Simulated acoustic fields showing a Gaussian beam passes through the metacage prototype. The incident wave travels from the left to the right. (c) and (d) show the simulated and measured acoustic fields in the scan area. Unit: cm.

The proposed metacage is also capable of shielding acoustic waves emitting from the interior, in which case the metacage acts as an acoustic enclosure with open channels. To demonstrate this, we place a loudspeaker at the center of the metacage with its front side facing up against the upper wall of the waveguide and measure the acoustic signal 1.5λ away from the outer surface of the metacage along a circle. The sound transmission loss is calculated with reference to the case where the metacage is removed. The directivity of the metacage with a 1.0 m/s air flow rate is also tested for comparison. An average sound transmission loss of over 10 dB is achieved between 2.3-2.5 kHz as shown in Fig. 5(a). Small variability of the transmitted acoustic pressure is observed when the metacage is rotated, which may be caused by the imperfection of the sample, directivity of the speaker, and measurement errors. The relatively small deviation demonstrates the omni-directivity of the acoustic metacage. As a comparison, we simulate the case of a point source placed inside the metacage, and the calculated energy decay with the metacage is plotted in Fig. 5(a). The average sound reduction over all angles in the simulation is about 14 dB at 2.5 kHz, slightly higher than that in the measurement. The corresponding acoustic field in simulation is depicted in Fig. 5(b) and some surface waves are clearly observable on the outer edge of the metacage. The far field sound field has very low amplitudes due to the sound insulation of the metacage. To show other situations where more than one point source is inside the metacage, different source pressure fields have been simulated and summarized in [33].

Finally, the air flow rate is measured when the metacage is present with the measurement setup shown in Fig. 5(c) to demonstrate the capability of allowing airflow. The metacage is covered with a plastic plate (not shown in the figure) to ensure airflow through the metacage only. Another measurement is performed where the metacage is absent and the locations of the wind speed meter and fan remain unchanged. A wind speed meter (type Holdpeak HP 866B) is placed outside the metacage to measure the air flow rate. The same fan (type FSY40S24M) used in the previous measurements is placed inside the metacage to generate airflow with the driving voltage being 27V. The measured air flow rates with and without the metacage are 0.3 m/s and 0.8 m/s, respectively. This measurement clearly demonstrates that the metacage is capable of allowing the exchange of airflow. The transmission of airflow can also be potentially increased by adjusting the sizes of the Helmholtz resonators so that the air channels can be wider.

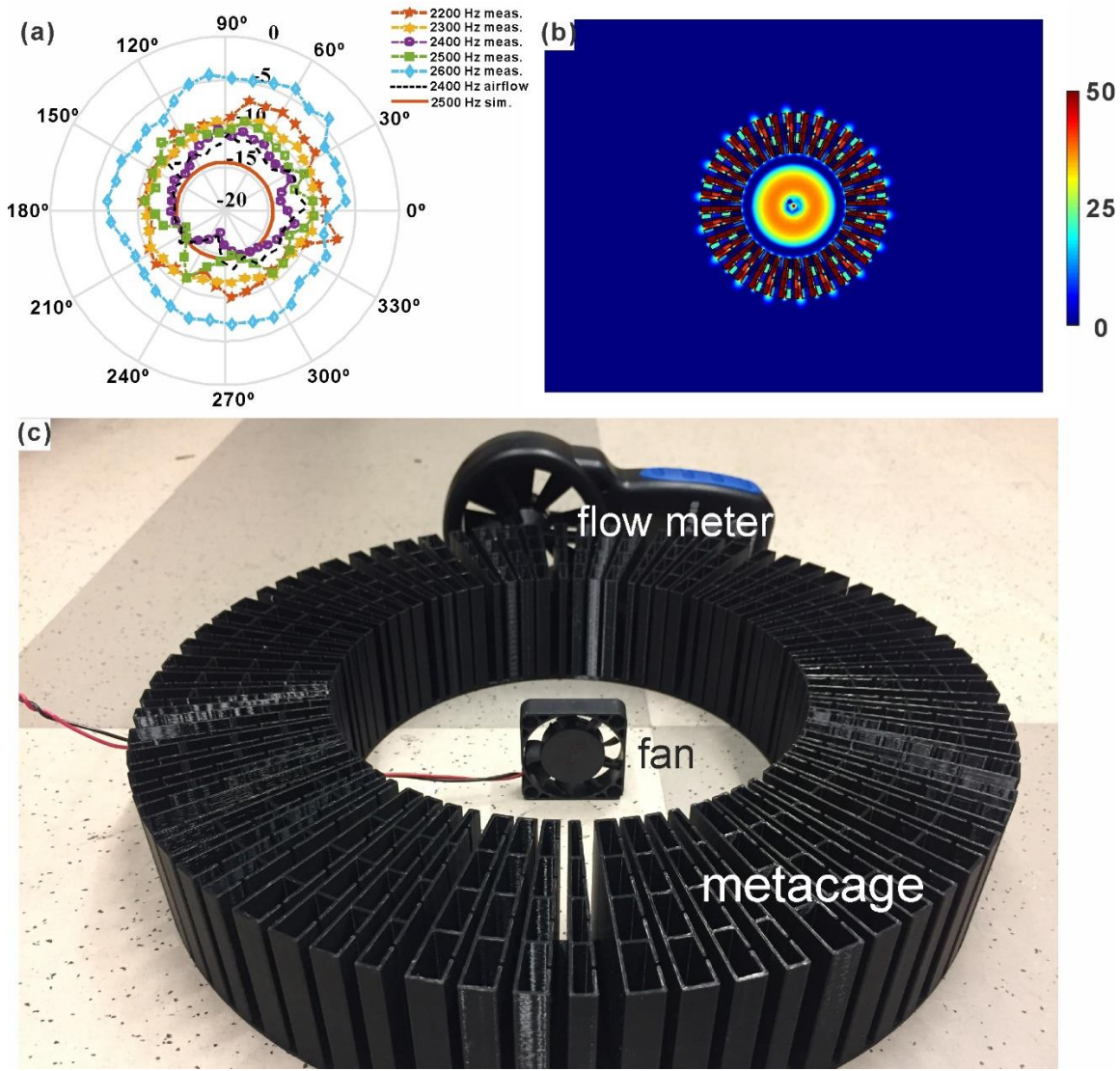

Fig. 5. Inside-to-outside performance of the proposed metacage. (a) Sound transmission loss through the metacage at different angles and frequencies. Unit: dB. (b) Simulated acoustic pressure field when the source is placed inside the metacage. Unit: dB. (c) Experiment setup of the flow rate measurement. Compared with the control case (no metacage, 0.8 m/s), about 40% of the airflow (0.3 m/s) can pass through the metacage.

## V. CONCLUSION

To conclude, an approach for conformal omnidirectional shielding of acoustic waves with open channels has been proposed. Theoretical analysis reveals that a properly designed GIM becomes acoustically opaque to in-plane incoming waves from all angles when its phase gradient is sufficiently large. An open channel 2D acoustic metacage based on this strategy is designed to prevent sound from passing through while allowing steady airflow, which can be important for sound insulation in ventilated environments. Numerical simulations and experiments demonstrate the effectiveness of the metacage. There are other existing structures in literature that may allow transport of airflow, such as phononic-crystal like pillars [42–44] and holey plates [45]. However, these have not been tested for airflow to the best of our knowledge. In addition, the metacage proposed here has a subwavelength thickness and could form a full enclosure, which makes it advantageous for certain applications. Although the working bandwidth is limited by using the current structure [30,38], it can be very useful for reducing tonal noises (such as those from various engines and fans) which sometimes can be more annoying than broadband noise [46]. Moreover, to deal with situations where noise of multiple frequencies is at present, multiple layers of the proposed metacages can be employed [33]. The realization of omnidirectional shielding of acoustic waves in such a compact and opened manner adds new capabilities for manipulating acoustic waves without impeding airflow. It is hoped that the design studied in this work can be helpful on the control of acoustic waves in various situations.

**Acknowledgement**

This work was supported by the Multidisciplinary University Research Initiative grant from the Office of Naval Research (N00014-13-1-0631) and NSF grant No. EFRI-1641084.